\documentclass{ifacconf}

\usepackage{graphicx}      
\usepackage{natbib}        

\usepackage[dvips]{epsfig}    

\usepackage{xcolor}

\usepackage{amssymb}
\usepackage{amsmath}          
\usepackage{enumerate}

\begin{document}
\begin{frontmatter}

\title{Robust asymptotic observer of motion states with nonlinear
friction}

\thanks[]{\textcolor[rgb]{0.00,0.00,1.00}{Author accepted manuscript (IFAC World Congress 2023)}}

\author{Michael Ruderman}
\address{University of Agder \\
Department of Engineering Sciences \\
P.B. 422, Kristiansand, 4604, Norway \\
email: \tt\small michael.ruderman@uia.no}

\begin{abstract}
This paper revisits the previously proposed linear asymptotic
observer of the motion state variables with nonlinear friction and
provides a robust design suitable for both, transient presliding
and steady-state sliding phases of the relative motion. The class
of motion systems with the only measurable output displacement is
considered. The reduced-order Luenberger-type observer is designed
based on the obtained simplified state-space representation with a
time-varying system matrix. The resulted observation error
dynamics proves to be robust and appropriate for all variations of
the system matrix, which are due to the nonlinear
spatially-varying friction. A specially designed tribological
setup to accurately monitor the relative motion between two
contacting friction surfaces is used to collect the experimental
data of the deceleration trajectories when excited by a series of
impulses. The performance of the state estimation using the
proposed observer is shown based on the collected experimental
data.
\end{abstract}


\end{frontmatter}


\section{Introduction}
\label{sec:1}

The observation of dynamic states in the motion systems is a
long-term problem within (motion) control communities. Often, only
the relative displacement of an actuated system is measured by the
sensing elements, such as e.g. encoders, while the relative
velocity as well as some internal dynamic states of the entire
system need also to be known in real time. The classical
observation strategies, like Luenberger observer (see
\cite{Luen1964}), disturbance observer (see e.g.
\cite{Ohishi87,Shim2016}), or sliding-mode observers (see e.g.
\cite{shtessel2014} and references therein), are well known. Also
a variety of related and analogous approaches have been developed
over the past three decades; several more specific references are
omitted here for reasons of space and too large a number published
works. One of the fundamental problems related to dynamic states
of relative motion and their observation is nonlinear friction,
which is present in almost all driven mechanical systems. As well
known, the kinetic friction can (i) lead to the undesired and
long-period stick-slip cycles in case of a feedback-controlled
motion (see e.g. \cite{ruderman2021}), (ii) affect the motion
breakaway and reversals phases (see e.g. \cite{ruderman2017} and
\cite{ruderman2017b}), and (iii) result in unacceptable
steady-state control errors (see e.g. \cite{ruderman2016}) if it
is not properly compensated. Moreover, the friction forces are
well known to be not directly measurable (see e.g.
\cite{harnoy2008}) and, in addition, they are mostly uncertain due
various system-internal and environment-external factors (see e.g.
\cite{ruderman2015}). A robust, like discontinuity-based,
compensation of nonlinear friction (see e.g. \cite{Ruder22}) can
partially solve the problem, but can have a limited acceptance in
several applications due to a high-frequent relay-type control
action. Therefore, it is often understood that an accurate on-line
estimation of friction force values can be of great benefit to the
performance of motion control systems with friction, cf. e.g.
\cite{olsson1998} and \cite{chen2000nonlinear}. Since the dynamics
of frictional forces are far from trivial, see e.g.
\cite{AlBender2008} for a detailed tutorial, designing a reliable
friction observer remains a significant challenge for the motion
control and its applications. Some recent experimentally confirmed
scenarios revealing the nature of frictional perturbations for
motion control can be found in the literature, for instance in
\cite{beerens2019,kim2019} just to name a few applications here.
An observer-based strategy for compensating the nonlinear friction
was proposed in \cite{ruderman2015}, which allowed for an accurate
positioning control up to the level of encoder resolution.
However, a systematic analysis of the error dynamics and, hence,
observer synthesis remained partially undisclosed and less
explained as methodology. The aim of this paper is to close this
gap and thus to approach the observer design for motion systems
with nonlinear friction in a simple and systematic way.

The rest of the paper is organized as follows. In the next
section, we first formalize the problem statement by introducing
the class of motion systems with nonlinear friction, for which the
asymptotic Luenberger-type state observer is proposed. In the main
section \ref{sec:2}, we make assumptions about the dynamic
friction state and introduce the corresponding asymptotic state
observer, also providing the robust design of observer gains. The
developed tribological experimental setup used in this work is
described in section \ref{sec:3}. Experimental evaluation of the
proposed observer is provided in section \ref{sec:4}. Conclusions
are given in section \ref{sec:5}.

\section{Problem formulation}
\label{sec:1a}

The following observation problem is addressed in this work.
Consider the dynamic motion system
\begin{equation}\label{eq:1:1}
    m \ddot{x}(t) + f\bigl(\dot{x}(t)\bigl) = u(t),
\end{equation}
where an inertial mass $m$ is actuated by the input $u$ and
counteracted by the nonlinear kinetic friction $f(\cdot)$. Recall
that the latter appears due to the normal contacts. The available
input and output signals are the drive force $u(t)$ and relative
displacement $x(t)$, both in the generalized coordinates of a 1DOF
system, here without distinguishing between the translational and
rotational motion. The inertial mass is assumed to be accurately
known, while for the nonlinear friction the following can be
assumed:
\begin{enumerate}[(i)]
    \item The total kinetic friction in steady-state is a superposition
    of the Coulomb and viscous friction forces
    \begin{equation}\label{eq:1:2}
    f(t) = F_c + F_v.
    \end{equation}
    The superposition \eqref{eq:1:2} is consistent with the established
    approaches to modeling friction, especially when using the Newton-Euler
    dynamics equations, and also considering the Stribeck weakening effects.
    The superposition principle can be temporary lost during the
    dynamic transients, particularly at the reversals of motion,
    where the viscous friction effects can subside and also lose their linearity.

    \item The Coulomb friction force in steady-state
    depends on the motion direction only, i.e.
    \begin{equation}\label{eq:1:3}
    F_c = C_f \, \mathrm{sign} ( \dot{x} ),
    \end{equation}
    and is parameterized by an uncertain Coulomb coefficient, for
    which a nominal value $C_f > 0$ is known.

    \item The viscous friction force depends linearly on the
    relative displacement rate only, i.e.
    \begin{equation}\label{eq:1:4}
    F_v = \sigma \, \dot{x}.
    \end{equation}
    The proportional friction law constitutes the
    standard linear system damping and is parameterized by the (generally) uncertain
    viscous friction coefficient, for which the nominal value $\sigma > 0$ is known.

    \item The transient behavior of kinetic friction $f(t)$ is to a large extent
    unknown, while assuming two of its commonly stated properties.
    (a) The viscous friction term is subject to the
    so-called frictional lag during sufficiently large velocity rates,
    i.e. at high accelerations and decelerations. (b) The Coulomb
    friction during the so-called \emph{presliding}\footnote{The term presliding
    is associated with a regime of friction where
    adhesive forces due to asperity contacts dominate, and thus the
    friction force is primarily a local function of relative
    displacement rather than velocity, see e.g. in \cite{armstrong1994,AlBender2008} for
    further details.}, i.e. at the motion beginning, stop, and
    reversals, is not discontinuous (cf. with eq. \eqref{eq:1:3})
    and undergoes some smooth hysteresis-shaped transitions.
    These transitions are subject to uncertainties due to the spatially
    and temporally varying nature of the friction surfaces.
\end{enumerate}

For the above system \eqref{eq:1:1}, with the nonlinear kinetic
friction satisfying (i)--(iv), one wishes to design a robust
asymptotic state observer so that the estimates of dynamic states
$\dot{\tilde{x}}(t) \rightarrow \dot{x}(t)$, $\tilde{f}(t)
\rightarrow f(t)$ as $t \rightarrow \infty$. The initial values of
the real system states $(\dot{x},f)(0)$ are unknown. And the only
available system measurements are $-\infty <\bigl(x(t), u(t)\bigr)
< \infty$, both affected by the sensing noise.

\section{State observer}
\label{sec:2}

\subsection{Dynamic friction states} \label{sec:2:sub:1}

At fist, we have to consider both frictional terms in
superposition, cf. eq. \eqref{eq:1:2}, as the dynamic friction
states. This will enable their observation and corresponding
integration into the observer scheme to be designed.

For the viscous friction term, one can write
\begin{equation}\label{eq:2:1:1}
\dot{F}_v = \beta^{-1} (\sigma \dot{x} - F_v),
\end{equation}
instead of \eqref{eq:1:4}, thus capturing the so-called frictional
lag, see e.g. \cite{AlBender2008} for details. Recall that the
latter appear (but not necessarily as linear) in the relative
$(\dot{x},F_v)$ coordinates. One can recognize that if the time
constant $\beta$ of the frictional lag is zero, then the dynamic
equation \eqref{eq:2:1:1} will collapse, and one will obtain the
standard viscous damping relationship which is static, cf. with
eq. \eqref{eq:1:4}. Here it is important to emphasize that the
introduced time constant $\beta > 0$ is relatively low and mostly
uncertain all the time. This makes its proper identification to a
sufficiently challenging task. If no nominal $\beta$ value is
available, assigning an arbitrary $0 < \beta \ll m\sigma^{-1}$ may
suffice. This will ensure the time constant of the friction lag is
significantly lower than the time constant of the mechanical
drive, cf. with eq. \eqref{eq:1:1}.

The dynamics of the nonlinear Coulomb friction captures the
presliding transition curves, this way leading to
\begin{equation}\label{eq:2:1:2}
\dot{F}_c = \left\{%
\begin{array}{ll}
    \dot{x} \cdot \partial F_c / \partial x \, , & \hbox{if } |F_c| < C_f,  \\[1mm]
    0, & \hbox{otherwise.} \\
\end{array}%
\right.
\end{equation}
Recall that describing the Coulomb friction force which should
include presliding and, therefore, avoid discontinuity at motion
reversals, relies on a smooth multi-valued mapping $x \mapsto f$
within the presliding range. The latter means that the motion
undergoes an onset or reversal, so that $|\dot{x}|$ is relatively
low and the non-viscous friction mechanisms predominate. As long
as a progressing $F_c(x)$-curve is not saturated at $\pm C_f$, the
dynamic transitions \eqref{eq:2:1:2} take place. Afterwards, the
constant Coulomb friction force characterizes the steady-state,
cf. with eq. \eqref{eq:1:3}.

Since both dynamic friction states \eqref{eq:2:1:1},
\eqref{eq:2:1:2} act on the motion dynamics \eqref{eq:1:1}
simultaneously, i.e. in superposition \eqref{eq:1:2}, their proper
decomposition can pose serious (probably unsolvable) challenge
when observing them individually. Therefore, the overall dynamic
friction state $f(\cdot)$ will be used next in the observer
design, while assuming that the nominal value $\sigma$ and
presliding map \eqref{eq:2:1:2} are available and the parameter
$\beta$ is sufficiently preestimated.

\subsection{Asymptotic Luenbeger-type observer} \label{sec:2:sub:2}

For the class of motion systems given by \eqref{eq:1:1}, with the
dynamic friction \eqref{eq:1:2}, we introduce the state vector $w
\equiv (w_1, w_2, w_3)^T : = (x, \dot{x}, f)^T$ and derive the
simplified dynamic state-space model as
\begin{equation}\label{eq:2:2:1}
\dot{w} = \underset{A} {\underbrace{
\left(%
\begin{array}{ccc}
0     & 1     & 0 \\
0     & 0     & -1/m \\
0     & \bigl(\partial F_c / \partial x +\sigma/\beta \bigr)  &  0
\end{array}%
\right) } } w + \underset{B} {\underbrace{
\left(%
\begin{array}{c}
  0 \\
  1/m \\
  0
\end{array}%
\right)
 } }
u.
\end{equation}
The associated output coupling vector is $C = (1,0,0)$, since the
relative displacement is the single available output value, i.e.
$x = C w$. One should notice that the system matrix $A$ has one
time-dependent term $\partial F_c / \partial x$ so that one deals
with $A(t)$ during the presliding. Outside the presliding range,
the $\partial F_c/
\partial x$ term becomes zero. On the contrary, $\partial F_c / \partial x \rightarrow \kappa$
after each reversal point $x(t_r^+) \equiv x_r$, where $t_r$ is
the time instant of the last motion reversal, and $\kappa \gg 0$
is a large positive constant representing the initial stiffness of
the given frictional surface pair. Note that one can have $\kappa
\rightarrow \infty$ at $t_r$, that is in line with temporary
stiction during the motion reversals, cf. \cite{ruderman2017b}.
Also recall that at $t_r$ the sign of $\dot{x}$ changes, i.e. the
system \eqref{eq:1:1} either starts or stops to move, or it
changes the direction of relative displacement. For both of the
above cases, i.e. for $\partial F_c / \partial x = 0$ and
$\partial F_c /
\partial x \gg 0$, the $(A,C)$-pair proves to be observable in
the Kalman sense. This allows designing the standard
Luenberger-type observer, cf. \cite{Luen1964}, so that an
asymptotic convergence of the states estimation, i.e.
$\tilde{w}(t) \rightarrow w(t)$, can be guaranteed.

In order to simplify the observer design and, moreover, to improve
the convergence properties of $\tilde{w}(t)$, we transform the
state-space representation \eqref{eq:2:2:1} into the regular form
\begin{eqnarray}
\label{eq:2:2:2}
\left(%
\begin{array}{c}
  \dot{\bar{w}} \\
  \dot{z} \\
\end{array}%
\right) &=&
\left(%
\begin{array}{cc}
A_{11}     & A_{12}  \\
A_{21}     & A_{22}
\end{array}%
\right) \left(%
\begin{array}{c}
  \bar{w} \\
  z \\
\end{array}%
\right) + \left(%
\begin{array}{c}
  B_{\bar{w}} \\
  B_z \\
\end{array}%
\right) \, u
 \\
  x &=&   \left(%
\begin{array}{cc}
  C_{\bar{w}} & C_z \\
\end{array}%
\right)  \left(%
\begin{array}{c}
  \bar{w} \\
  z \\
\end{array}%
\right),
  \label{eq:2:2:3}
\end{eqnarray}
where $\bar{w} = w_1$ and $z=(w_2,w_3)^T$. Then, we apply a
standard reduced-order asymptotic observer, cf.
\cite{luenberger1971}. Recall that the above transformation of
\eqref{eq:2:2:1} into the regular form, with breakdown of $A$,
$B$, $C$ into the matrices of an appropriate dimension in
\eqref{eq:2:2:2}, \eqref{eq:2:2:3}, provides a separation into the
measurable and unmeasurable states $\bar{w}$ and $z$,
respectively. Subsequently, an exclusion of $\bar{w}$ from the
estimated state vector reduces the order of the observer dynamics
from three to two and, this way, the complexity of the associated
poles assignment.

The reduced-order Luenberger observer is given by
\begin{eqnarray}
\label{eq:2:2:4}
\dot{\tilde{z}} &=& (A_{22}-LA_{12}) \tilde{z} + \\
\nonumber   &+& (A_{21} - L A_{11} + A_{22}L -LA_{12}L) \bar{w} +
(B_z -L B_{\bar{w}}) u,
\end{eqnarray}
where $L \equiv (L_1, L_2)^T$ is the vector of observer gains,
which are the design parameters. Here $\tilde{z}$ is the estimate
vector of the unmeasurable system states $z$. It is worth
recalling that the state observer \eqref{eq:2:2:4} provides an
asymptotic convergence $\tilde{z}(t) \rightarrow z(t)$ under one
and the single condition -- the observer system matrix
$(A_{22}-LA_{12})$ has to be Hurwitz. In order to transform back
the dynamic state $\tilde{z}(t)$, which was input-output-scaled
according to the right-hand-side of \eqref{eq:2:2:4}, the
back-transformation
\begin{equation}\label{eq:2:2:5}
(\tilde{w}_2, \tilde{w}_3)^T = \tilde{z} + L \bar{w},
\end{equation}
is subsequently required, cf. \cite{luenberger1971}.

\subsection{Presliding transitions} \label{sec:2:sub:3}

\begin{figure}[!h]
\centering
\includegraphics[width=0.7\columnwidth]{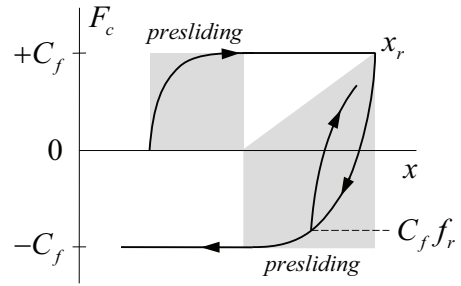}
\caption{Presliding friction force transitions at reversals. Each
change in the sign of $\dot{x}$ gives rise to a new hysteresis
branch and sets the frictional system into presliding. Once the
friction force saturates at $\pm C_f$, the hysteresis memory is
erased, and the frictional system is in gross sliding.}
\label{fig:1}
\end{figure}

In order to capture the $\partial F_c / \partial x$ term, which is
entering the system matrix $A$ in \eqref{eq:2:2:1}, one needs to
map the continuous presliding friction transitions in the sense of
$\bigl(x(t)-x_r\bigr) \mapsto F_c$. Note that a particular
modeling approach can be of secondary importance when capturing
the presliding friction force. Most important is rather to ensure
that such mapping is (i) piecewise continuous on the interval
between the reversal state $f_r \equiv F_c(t_r)$ at $t_r$ and the
saturated friction state $|F_c| = C_f$; (ii) the
force-displacement curves are at least $\mathcal{C}^1$ smooth,
(iii) the force-displacement curves follow the closed hysteresis
loops between two reversal points and, then, they saturate at the
$\pm C_f$ Coulomb friction level, cf. Fig. \ref{fig:1}.
While different fairly suitable modeling approaches (most
importantly, those that have a small number of free parameters)
exist, for instance the Dahl model (see \cite{dahl1976}) or the
modified Maxwell-slip model (see \cite{ruderman2011}), in this
work we use an approach (see \cite{ruderman2017} for details)
which is based on the tribological study \cite{koizumi1984} of
presliding hysteresis loops. Using the scaling factor $s > 0$,
which relates an after-reversal motion to the presliding distance
as
\begin{equation}\label{eq:2:3:1}
z = s \int \limits^{t}_{t_{r}} \dot{x} \, dt,
\end{equation}
and is defined on the interval $z \in [-1, \, 0) \cup (0, \, 1]$,
one can describe the branching of friction force in presliding by
\begin{equation}\label{eq:2:3:2}
f_0(z)= z \bigl(1-\ln|z|\bigr).
\end{equation}
Recall that each motion reversal at $t_r$ gives rise to a new
presliding transition, so that the total (normalized) presliding
friction map is, cf. \cite{ruderman2017},
\begin{equation}\label{eq:2:3:3}
f_p(t)= \bigl| \mathrm{sign}(\dot{x}) - f_r \bigr| \, z
\bigl(1-\ln|z|\bigr) + f_r.
\end{equation}
Note that the presliding friction dynamics memories the state of
the last reversal transition, i.e. $f_r := f_p(t_r)$, see Fig.
\ref{fig:1}. Since the presliding mapping \eqref{eq:2:3:3} is
defined for $|z| \leq 1$ only, the entire continuous Coulomb
friction law becomes, cf. with eq. \eqref{eq:1:3},
\begin{equation}\label{eq:2:3:4}
F_c= \left\{%
\begin{array}{ll}
    C_f \, f_p(t) , & \hbox{if } |z|
\leq 1,  \\[1mm]
    C_f \, \mathrm{sign}\bigl(\dot{x}(t)\bigr), & \hbox{otherwise.} \\
\end{array}%
\right.
\end{equation}
For the presliding distance $z$, which is linked to the output
displacement by the scaling parameter $s$ and integral
\eqref{eq:2:3:1}, which is reset at each $t_r$, one can obtain
$\partial F_c / \partial x$ out from
\begin{equation}\label{eq:2:3:4}
\frac{\partial F_c}{\partial z} = - C_f
\bigl|\mathrm{sign}(\dot{x})-f_r \bigr| \, \ln|z|.
\end{equation}

\subsection{Robust observer design} \label{sec:2:sub:4}

Recalling the time-variance of the system matrix $A$ and $A_{22}$,
respectively, one can show the characteristic polynomial of the
matrix $(A_{22}-LA_{12})$ to be
\begin{equation}\label{eq:2:4:1}
\Bigl(\lambda+L_1\Bigr)\lambda + \frac{1}{m} \Bigl( \sigma/\beta +
\partial F_c /
\partial x - L_2 \Bigr) = 0.
\end{equation}
Obviously $\lambda$ is the complex Laplace variable so that both
eigenvalues of \eqref{eq:2:4:1} can be computed explicitly as
\begin{equation}\label{eq:2:4:2}
\lambda_{1,2} = \frac{1}{2} \Bigl(-L_1 \pm \sqrt{L_1^2 +
\frac{4}{m}\bigl( L_2 - \partial F_c / \partial x - \sigma/\beta
\bigr)} \, \Bigr).
\end{equation}
Based on that, imposing the conditions
\begin{equation}\label{eq:2:4:3}
(\mathrm{a}) \quad L_1 > 0 \quad \text{ and } \quad (\mathrm{b})
\quad L_1
> 2\sqrt{\frac{\kappa+\sigma/\beta-L_2}{m}}
\end{equation}
one can ensure the observer dynamics is (i) asymptotically stable
and (ii) has no complex poles, which otherwise would lead to
transient oscillations. Furthermore, one can recognize that a
sufficiently large $L_1$ will determine the most left-hand-side
pole and, thus, a faster convergence of the $\tilde{w}_2$
estimate, which is the relative velocity. At the same time, $L_2$
is expected to be negative in the most cases, i.e. for $0 < \sigma
\ll \kappa$, so as to guarantee that \eqref{eq:2:4:3} holds also
for $\partial F_c /
\partial x = 0$. Recall that the latter characterizes the sliding
phases where $F_c$ is already saturated. A selected ratio between
the $L_1, L_2$ gains, while satisfying \eqref{eq:2:4:3}, is
controlling the distance between both real poles \eqref{eq:2:4:2}.
This distance becomes maximal for $\partial F_c /
\partial x = 0$, and it is decreasing during the presliding as the
$\partial F_c / \partial x$ value grows. This way, a robust
observer \eqref{eq:2:2:4}, \eqref{eq:2:2:5} can be realized for
all phases of the nonlinear friction, also for uncertain
parameters and during transient behavior of the dynamic friction,
cf. sections \ref{sec:1} and \ref{sec:2:sub:1}.

\section{Tribological setup}
\label{sec:3}

\begin{figure}[!h]
\centering
\includegraphics[width=0.95\columnwidth]{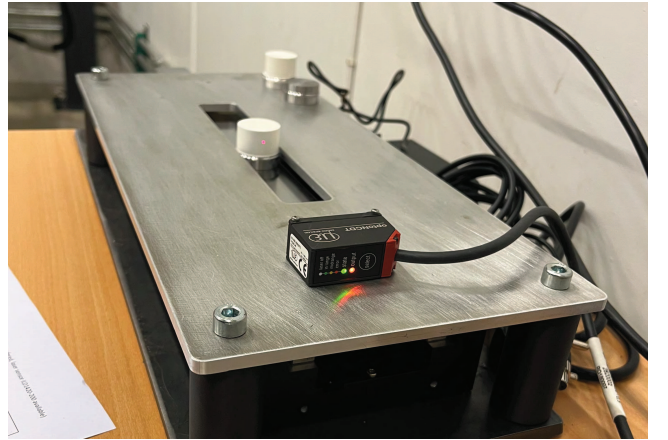}
\caption{Tribological experimental setup (laboratory view).}
\label{fig:3:1}
\end{figure}

The specially designed tribological setup (see Fig. \ref{fig:3:1})
consists of a linear moving platform which is placed under the
mechanical frame and guides the lumped disks (from various
materials). This way, an exposed disk and the linear platform have
a homogenous horizontal frictional contact. The used steel disk
(lacquered for a better laser reflection) with the total mass
$m=52$ gram can slide over the moving platform which has an
equally polished surface out of steel. The Coulomb friction
coefficient is calculated as $C_f = \mu m g = 0.2143$ N, where $g$
is the gravitational acceleration constant. The frictional
coefficient $\mu$ is taken from the tribological literature for a
kinetic (sliding) behavior of a clean and dry steel on steel pair.
The moving platform can be actuated by a servo-drive with the
stiff high-precision ball-screw, so that its linear displacement
is known based on the motor encoder (not used in this work). The
absolute position of the sliding disk is measured by a
high-resolution laser sensor, which beam is touching the disk in
its center (see Fig. \ref{fig:3:1}). The diameter of the disk is
only half millimeter lower than the width of the guiding slot of
the frame. This way, the motion of the disk is accurately
constrained from both sides, and the only one horizontal degree of
freedom (i.e. denoted with the coordinate $x$) can be assumed.
Lateral contacts between the disc and the walls of the frame-slot
are minimized by the specially machined side-edging on the disc.
Therefore, the by-effect due to the side-contact with the walls
can be neglected, in comparison to the main frictional contact
area of the disc which is placed on the moving platform.

In this work, the decelerating trajectories of the sliding disk
were used without actuating the motion platform. More
specifically, a series of mechanical impulses was manually
provided to the disk, while the motion platform remained
non-driven. This way, the measured with the laser sensor signal
$x(t)$ represents an absolute displacement of the disk, while an
initial relative velocity after each new impulse is unknown. Note
that the unfiltered real-time data of the disk displacement are
recorded in the control board, with the sampling rate set to 2
kHz.

\section{Experimental evaluation}
\label{sec:4}

\begin{figure}[!h]
\centering
\includegraphics[width=0.99\columnwidth]{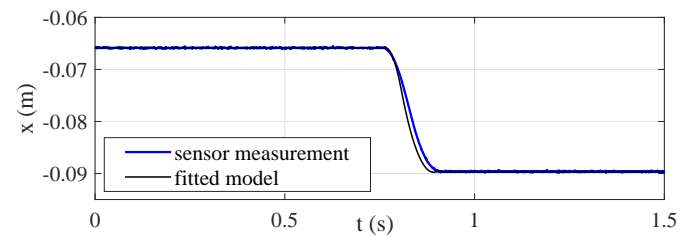}
\caption{Measured impulse response versus the motion dynamics
\eqref{eq:1:1} with least-squares best fit of friction
\eqref{eq:1:2}.} \label{fig:4:1}
\end{figure}

\begin{figure}[!h]
\centering
\includegraphics[width=0.99\columnwidth]{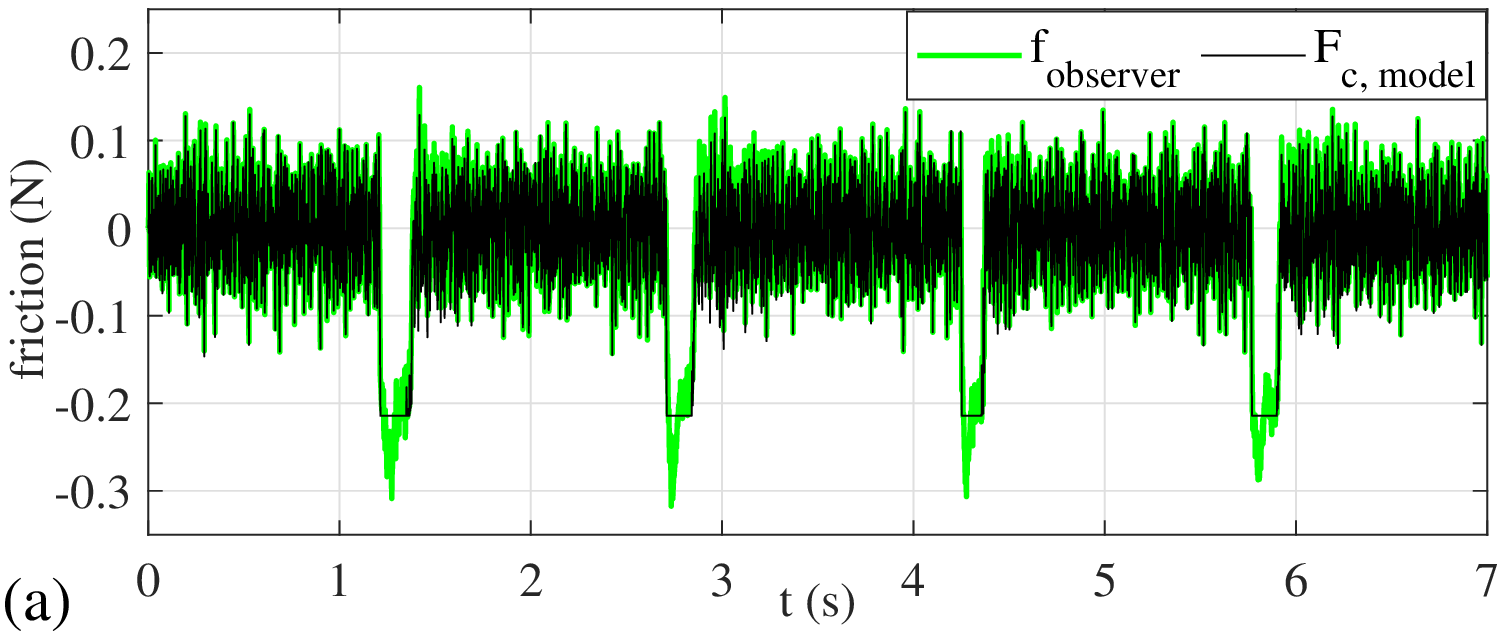}
\includegraphics[width=0.99\columnwidth]{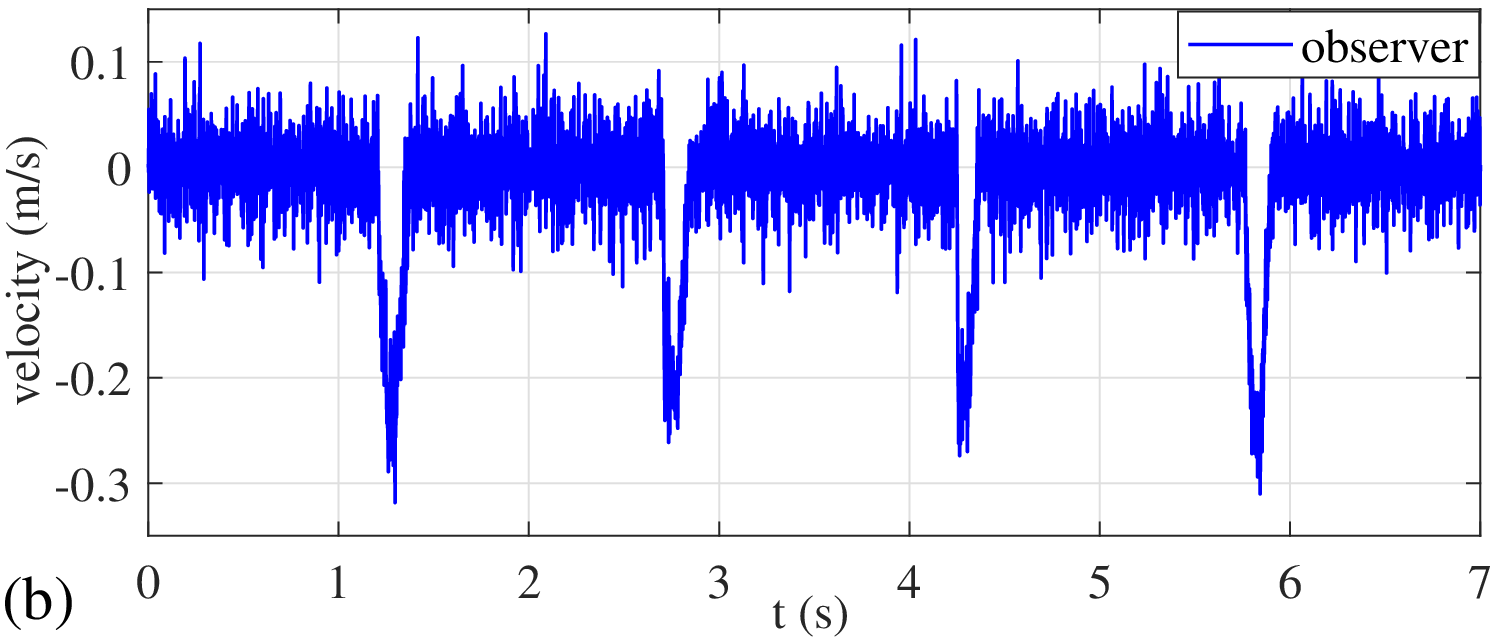}
\includegraphics[width=0.99\columnwidth]{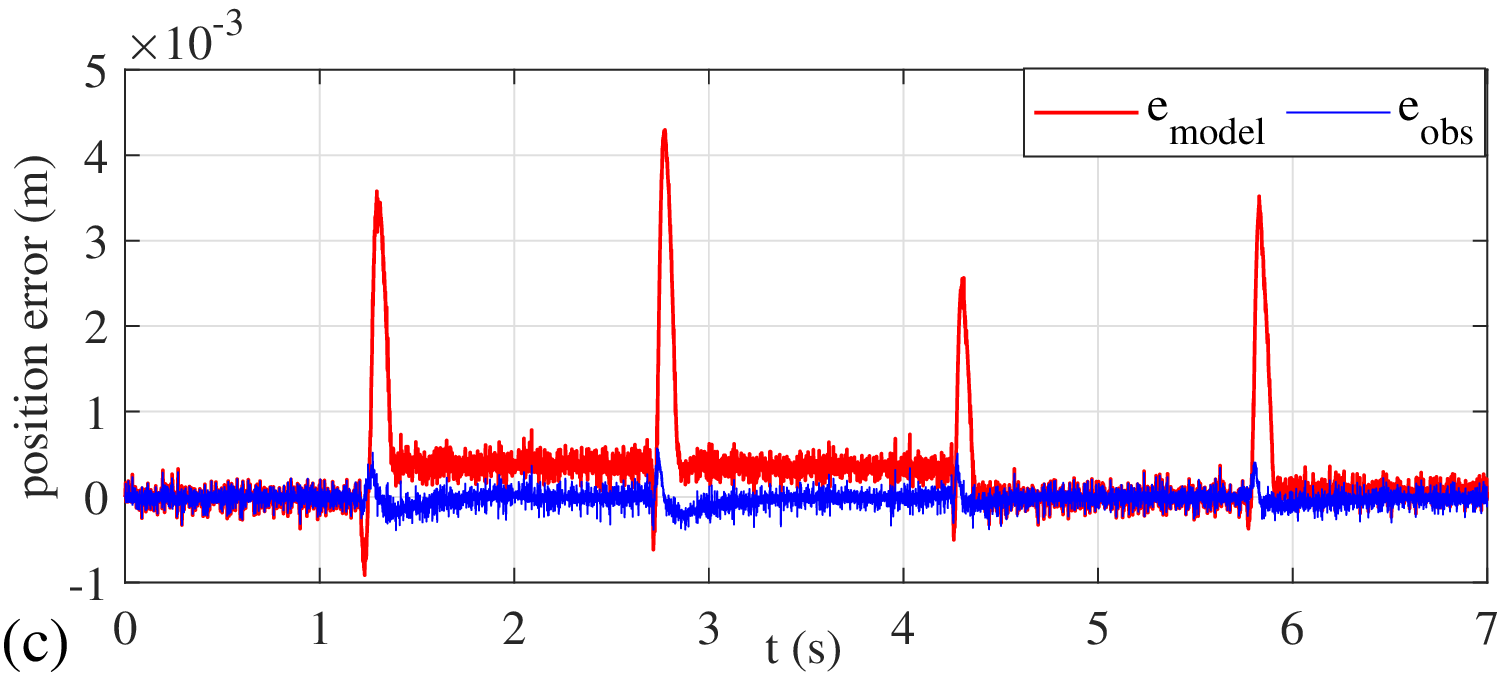}
\caption{Observer evaluation: model-predicted $F_c$ versus
observed $\tilde{w}_3$ (a), observed $\tilde{w}_2$ (b),
model-predicted position error $e_{model}$ versus observer error
$e_{obs}$ (c).} \label{fig:5}
\end{figure}

While the overall moving mass $m$ and the Coulomb friction
coefficient $C_f$ are assumed to be known from the available
system data, cf. section \ref{sec:3}, the residual frictional
parameters $\sigma$, $\beta$, $s$ are uncertain and can only be
estimated roughly, based on the (even though accurate)
experimental measurements. The least-squares best fit of the
modeled dynamics \eqref{eq:1:1} when tuning the unknown frictional
parameters in \eqref{eq:1:4}, \eqref{eq:2:1:1}, \eqref{eq:2:3:1},
is shown in Fig. \ref{fig:4:1}.
Note that the manually injected input impulse cannot be ideally
executed, i.e. with a time-of-impact $\rightarrow 0$. Therefore,
the measured impulse response is inherently affected by some
by-effecting transient input dynamics, cf. Fig. \ref{fig:4:1},
even though the impulse is also fitted in the model together with
the identified parameters. Despite some visible transient
discrepancy, the principal shape of the fictionally damped
convergence is well captured by the modeled $f(\cdot)$, cf. later
with Fig. \ref{fig:5} (c). This allows implementing the
state-space model \eqref{eq:2:2:1} and the asymptotic observer
\eqref{eq:2:2:4}, \eqref{eq:2:2:5}.

\begin{figure}[!h]
\centering
\includegraphics[width=0.99\columnwidth]{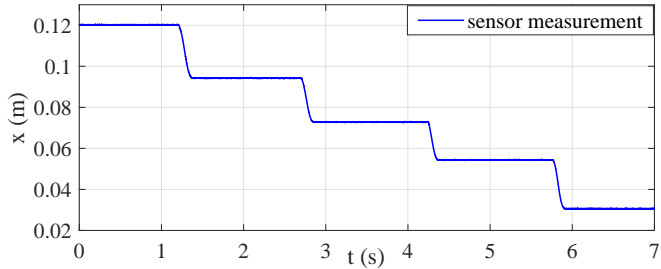}
\caption{Relative motion of the sliding disk: measured
displacement in response to the series of impulses.}
\label{fig:4:2}
\end{figure}

The measured motion profile, shown in Fig. \ref{fig:4:2} for a
sequence of manually injected mechanical impulses, is used for
evaluating the designed observer.
Note that the applied impulses provide a series of unknown
disturbing initial conditions, for which the robust observer has
to converge fast enough during the relatively short periods of
non-zero velocity, cf. Fig. \ref{fig:4:2}. It is also important to
notice that the excited motion steps are distributed over the
entire length of the sliding surface of the supporting platform.
This results in the varying, correspondingly uncertain, parameters
of the frictional behavior, depending on spatial properties of the
contact surface.

The assigned poles of the observer are $\lambda_{1,2} = (-350,
-10)$. These result in the feedback gain values $L^T=(360, -182)$.
The evaluated observer performance is demonstrated in Fig.
\ref{fig:5}. The $\tilde{w}_3$-estimate, which is the observed
dynamic friction state, is compared with the model-predicted
Coulomb friction force in the diagram (a). Note that the
high-frequency pattern below the $C_f$-level is not discontinuous,
and it constitutes a smooth pres-sliding friction behavior driven
by the noisy $\bar{w}(t)$ and $\tilde{w}_2(t)$ values. The
corresponding $\tilde{w}_2$-estimate of the relative velocity is
shown in the diagram (b), cf. with the measured motion profile
from Fig. \ref{fig:4:2}. The estimated velocity peaks are clearly
visible despite the noise of displacement measurement is
propagated into the $\tilde{w}_2(t)$ signal. For a further
assessment of the observer performance, the position error is
compared between the nominal model, i.e.
$$
e_{model} = \bar{w} - x,
$$
where $x$ is predicted based on \eqref{eq:1:1}, and the
observer-based
$$
e_{obs} = \bar{w} - \int \tilde{w}_2 \, dt,
$$
where $\tilde{w}_2$ is the velocity estimate. Despite the
integrative behavior, the $e_{obs}(t)$ is clearly superior
comparing to $e_{model}(t)$, especially during the transients of
non-zero velocities, as it is visible in the diagram (c). This
speaks further in favor of the $(\tilde{w}_2, \tilde{w}_3)(t)$
estimate.

\section{Conclusions}
\label{sec:5}

A simple and robust asymptotic observer of the nonlinear friction
state is provided. Using the derived time-varying state-space
notation of the motion dynamics and reduced-order Luenberger
observer, it is shown that both dynamic states of the relative
velocity and overall nonlinear friction force can be estimated
sufficiently fast and accurately. In particular, the estimation is
feasible for both, the transient presliding and sliding phases of
an excited and damped relative motion. The observer provided can
be parameterized by a simple and straightforward pole placement
manner, which makes it well accessible for standard control
engineering applications. An experimental evaluation was
demonstrated on a specially designed tribological setup of moving
bodies on a sliding surface. The relatively short motion
instances, excited by a series of the external force-impulses,
where used as a challenging scenario of relative motion in the
transient phase.

\section*{Acknowledgment}
The author is grateful to the laboratory engineering support by
Roy Werner Folgero and Harald Sauvik for assembling the
tribological setup.

\bibliographystyle{alpha}        
\bibliography{references}             

\end{document}